\documentclass[twocolumn,showpacs,preprintnumbers]{revtex4}
\usepackage{amsmath}
\usepackage{epsfig}
\usepackage{graphicx}
\DeclareGraphicsExtensions{.eps}

\usepackage{amsmath}
\begin{document} 
\title{Bose gas with generalized dispersion relation plus an energy gap} 
\author{J. G. Mart\'inez-Herrera$^{1,2}$, J. Garc\'ia$^{1,2}$ and M. A. Sol\'is$^3$ \\ 
   {$^{1}$Posgrado en Ciencias F\'isicas, UNAM, 04510 Cd. Mx., M\'exico}\\
   {$^{2}$Instituto de F\'isica, UNAM, Apdo. Postal 20-364, 01000 Cd. Mx., 
     M\'exico}\\
}

\begin{abstract}
	Bose-Einstein condensation in a Bose gas is studied analytically, in any positive dimensionality ($d>0$) for identical bosons with any  energy-momentum positive-exponent ($s>0$) plus an energy gap $\Delta$ between the ground state energy $\varepsilon_0$ and the first excited state, i.e.,  $\varepsilon=\varepsilon_0$ for $k=0$ and $\varepsilon=\varepsilon_0 +\Delta+ c_sk^s$, for $k>0$, where $\hbar \mathbf{k}$ is the particle momentum and $c_s$ a constant with dimensions of energy multiplied by a length to the power $s > 0$. Explicit formula with arbitrary $d/s$ and $\Delta$ are obtained and discussed for the critical temperature and
	the condensed fraction, as well as for the equation of state from where we deduce a generalized $\Delta$ independent thermal de Broglie wavelength. Also the internal energy is calculated from where we obtain the isochoric specific heat and its jump at $T_c$. When  $\Delta > 0$, for any $d > 0$ exists a Bose-Einstein critical temperature $T_c \neq 0$ where the internal energy shows a peak and the specific heat shows a jump. Both the critical temperature and the specific heat jump increase as functions of the gap but they decrease as functions of $d/s$. 
	At sufficiently high temperatures $\Delta$-independent classical results are recovered while for temperatures below the critical one  the gap effects are predominant. For $\Delta = 0$ we recover previous reported results.
	
\end{abstract}
\date{Modified: \today / Compiled: \today}
\maketitle

\section{Introduction}

Immediately after the discovery of helium superfluidity by Kapitza \cite{Kapitza} and Allen \cite{Allen} independently, London \cite{London} proposed that this new phenomenon could be a consequence of the Bose Einstein condensation (BEC)  of the helium atoms. London himself calculated the BEC  critical temperature of a 3D ideal gas of helium atoms getting $T_c = 3.09$ K, which is really close to the $\lambda$-transition temperature 2.18 K. However, he was unable to reproduce the low temperature exponential behavior of the specific heat of Helium-4  ($^4$He) previously measured by Keesom \cite{Keesom}. In order to reproduce this low temperature experimental behavior, London \cite{London} and others \cite{Bijl,Toda,Matsubara} proposed an energy gap additional to the ideal gas energy spectrum which reproduced 
 the low temperature exponential behavior of the specific heat but at the same time this new energy spectrum also increases the BEC critical temperature away from the $\lambda$-transition temperature. Thenceforth, there were several theoretical efforts to justify this energy gap with the argument that it could be an effect of the interaction among the bosons \cite{Girardeau,Hugenholtz}, which was ruled out by Hugenholtz and Pines \cite{Hugenholtz}.

On the other hand, although for a three-dimensional homogeneous ideal Bose gas (IBG) of permanent particles there is a BEC critical temperature $T_c$ different from zero such that for $T < T_c$ a large portion of all particles are on the ground state forming a \emph{Bose-Einstein condensate} \cite{einstein}, this phenomenon is not observed at temperatures different from zero for 2D or 1D Bose gases, whether spatially homogeneous \cite{Pathria} or within spatially periodic structures (multilayeres, multitubes, ...), as can be deduced from results reported in Refs. \onlinecite{PatyPRA2010} and \onlinecite{PatyJLTP2012}, where external periodic potentials acting on a 3D IBG always decresed its BEC critical temperature below that of a free IBG.
Also, it is well known that external uniform potentials \cite{GretherEPJD2003,BagnatoPRA1987,BagnatoPRA1991,GreinerAPB2001,GreinerPhD2003} can change the particle energy-momentum relation of an IBG and, as a consequence, the dimensionality where BEC can exist as well as the magnitude of its critical temperature.

Recently, an interactionless Bose gas within {\it imperfect} periodic structures enclosed in an infinite box,  has showed \cite{ViridianaIJMPB2016} to have an increase in its BEC critical temperature in 3D multilayers as well as the  emergence of a BEC at temperature different from zero in 2D multisquares or 1D multirods \cite{G-J2017}; in all three cases, the increase in the BEC critical temperature with respect to the case of a free ideal Bose gas, is  caused by the appearance of an energy gap in the bosonic energy spectrum. 
Also, it has been possible to observe the emergence of an energy gap in the energy spectrum of interactionless particles promoted by the asymmetry of the complex network structure where they are. For example, the birth of an energy gap between the ground and the first excited states for bosons confined has been reported on: a discrete network with inhomogeneous topology such as the comb lattice \cite{BurioniEL2000} or the star graph \cite{BurioniJPB2001} where authors claim BEC occurs in a spectral dimension less than two; a star-shaped optical network \cite{Brunelli}; an infinitely ramified wheel graph and infinitely ramified star graph \cite{VidalPRE2011}; networks composed of interconnected linear chains \cite{BuonsantePRB2002}; or an Apollonian network \cite{OliveiraPRE2013}, where all of them present BEC at non-zero temperature due to the energy gap emergence.

In this work we generalize the London's toy model \cite{London} in such a way to report the energy gap effects on the magnitude and existence of the BEC critical temperature 
in any positive dimensionality ($d>0$) for a gas of zero-spin  identical bosons with an energy-momentum positive-exponent ($s>0$) plus an energy gap between the ground and the first excited states. Additionally, we give gap dependent generalized expressions for a few representative thermodynamic properties.  
%
%
In Sec. II we expose our system of interest and we derive a general expression for the BEC critical temperature as a function of the energy gap magnitude, the system dimensionality and the energy-momentum positive-exponent, from where we are able to claim the existence of BEC in any dimensions at critical temperature different from zero. In Sec. III we obtain the energy gap dependence for the equation of state from where we deduce a generalized thermal de Broglie wavelength, internal energy, isochoric specific heat as well as its
 jump at $T_c$. In Sec. IV we give our conclusions. 

\section{Bose gas with a generalized dispersion relation plus an energy gap, $s>$0, $d>$0, $\Delta\neq$0}

\begin{figure}
	\centering
	\centerline{\epsfig{file=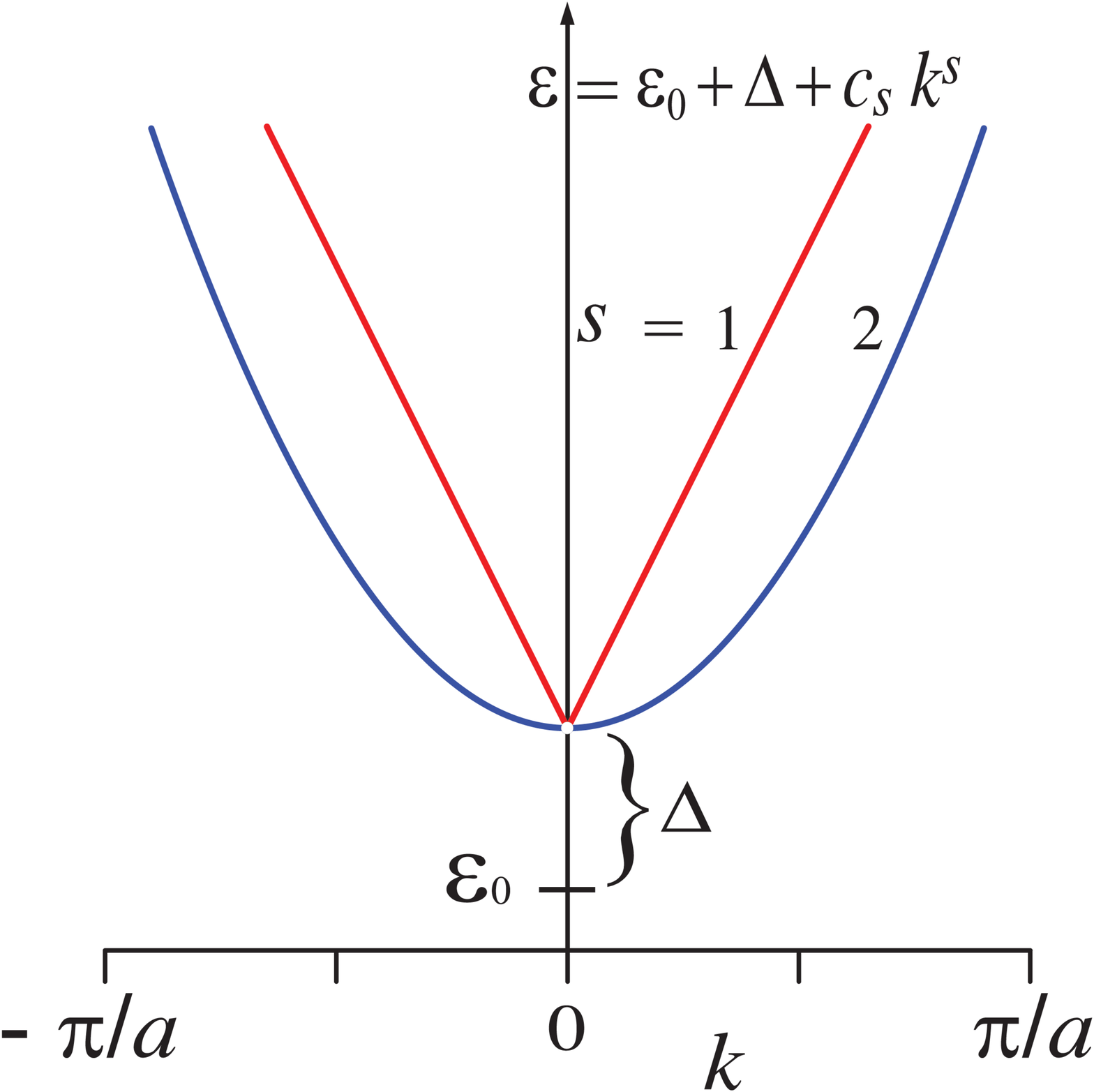,height=2.6in,width=3.0in}}
	\caption{(Color online) Linear and quadratic dispersion relations plus an energy gap.}
	\label{graf:relacion-dispersion}
\end{figure}

Our system is a $d$-dimensional infinite Bose gas 
of $N$ permanent identical zero-spin particles of mass $m$ and wave-vector magnitude $k$ whose particle energy is
\begin{equation}\label{eq:reldis}
\varepsilon=
\left\{
\begin{array}{ll}
\varepsilon_0+\Delta+c_s k^s,   & k > 0 \\
\varepsilon_0, & k = 0
\end{array}
\right.
\end{equation} 
where $\varepsilon_0$ is the ground state energy and $\Delta$ the particle energy gap needed 
to reach the first excited energy state, $c_s$ is a constant with dimension of energy multiplied by a length to the  power $s$ which is the exponent of $k$ in the dispersion relation. 
In Fig. \ref{graf:relacion-dispersion} we show two examples of dispersion relation (linear $s = 1$ and quadratic $s = 2$) plus an energy gap. 
\subsection{Critical Temperature} 
To calculate the critical temperature we begin with the particle number equation
\begin{equation}
N=N_0 + \sum_{\boldsymbol{k}\neq 0} n_k
\end{equation}
where $n_k=1/(e^{\beta(\varepsilon-\mu)}-1)$ is the Bose-Einstein distribution function, $N_0$ is the number of particles in the ground state, $\boldsymbol{k}$ is the particle wave-vector and $\mu$ is the chemical potential.

Taking the thermodynamic limit (T.L.) $N\rightarrow \infty$, and the system volume $L^d \rightarrow \infty$ but the number density $n\equiv N/L^d$ is a constant, $2\pi / L$ is infinitesimally small and the available states for a particle form a continuum.
Then we can change the sum by an integral over the entire $d$-dimensional $k-$space $\Omega$ in the following way
$$\sum_{\boldsymbol{k\neq0}}\rightarrow \left(\frac{L}{2\pi}\right)^d\int_{\Omega} d\boldsymbol{k}.$$ 
Thus we can write the number of bosons $N$ as the sum of the $N_0$ particles that are in the lowest energy level ($k=0$), plus the particles in the excited states
\begin{equation}
N=N_0 +\left(\frac{L}{2\pi}\right)^d\int_{\small{\Omega}} \frac{d\boldsymbol{k}}{e^{\beta(\varepsilon-\mu)}-1}.
\end{equation}
Integrating over the $d$-dimensional sphere, using the dispersion relation \eqref{eq:reldis} and changing the variable $k$ by $\varepsilon$ in the integral we get
\begin{equation}\label{num_bos_e}
N=N_0+\left(\frac{L}{2\pi}\right)^d \frac{2\pi^{d/2} (k_BT)^{d/s}}{sc_s^{d/s}\Gamma(d/2)} \Gamma(d/s)g_{\frac{d}{s}}(z_1)
\end{equation}
where $g_{\frac{d}{s}}(z_1)$ is the function of Bose of order $d/s$ with argument $z_1\equiv e^{\beta(\mu-\varepsilon_0-\Delta)}$. Note that when $\varepsilon_0=0$ and $\Delta=0$, $z_1$ reduces to the well-known fugacity of the infinite ideal Bose gas.

At BEC critical temperature $T_c$ the chemical potential $\mu$ takes the value of ground state energy $\varepsilon_0$  and the fraction of particles $N_0/N$ in the ground state is nearly zero. Thus, evaluating Eq. (\ref{num_bos_e}) at $T_c$, we obtain
\begin{equation}\label{tem_critica}
T_c = \frac{c_s}{k_B}\left(\frac{s \, (2\pi)^{d} \,\Gamma(d/2) \, n}{2\pi^{d/2} \, \Gamma(d/s)g_{\frac{d}{s}}\left(e^{-\frac{\Delta}{k_BT_c}}\right)}\right)^{s/d}.
\end{equation} 
From Eq. (\ref{tem_critica}) we can realize that the critical temperature is ground state energy independent 
 since $\varepsilon_0$ acts only as a reference energy. Moreover, we note that any finite energy gap $\Delta$ produces a critical temperature $T_c$ different from zero for $d/s>0$, while that, when $\Delta=0$, exists $T_c\neq0$ only if $d/s>1$ \cite{Aguilera}.
At this point we introduce a temperature unit $T_{0c}$ as the critical temperature of that Bose gas whose energy gap magnitude $\Delta_0$ is equal to $k_BT_{0c}$. This temperature unit allows us to analyze Bose gases even with $0<d/s\leq 1$, unlike other ones like the temperature defined as the critical temperature of the same gas but with energy gap $\Delta=0$ which is zero for $0<d/s\leq 1$. Note that $T_{0c}$ is $d/s$ dependent. From Eq. (\ref{tem_critica}) we obtain
\begin{equation}\label{eqtem_toc}
T_{0c}=\frac{c_s}{k_B}\left( \frac{s(2\pi)^d\Gamma(d/2) \, n}{2 \pi^{d/2} \Gamma(d/s) \, g_{\frac{d}{s}} \left(e^{-1}\right)}\right)^{s/d}
\end{equation}
where we have introduced the energy unit $k_B T_{0c}$.
Making the quotient between Eq. (\ref{tem_critica}) and Eq. (\ref{eqtem_toc}) we obtain the following dimensionless formula which is valid for any value of $d/s > 0$
\begin{equation}\label{fracción_Tc_T0c}
\vspace{-0.1cm}
\frac{T_c}{T_{0c}}=\left(\frac{g_{\frac{d}{s}}\left( e^{-1} \right)}{g_{\frac{d}{s}}\left( e^{-\frac{\Delta}{k_B T_{0c}}\frac{T_{0c}}{T_c}}\right)}\right)^{s/d}.
\end{equation}
Thus we can visualize how the temperature changes as a function of the energy gap and as a function of the ratio $d/s$ even for the region $0<d/s \leq 1$.
%
\begin{figure}[htb]
\centering
\hspace{1cm}
\includegraphics[scale=0.4]{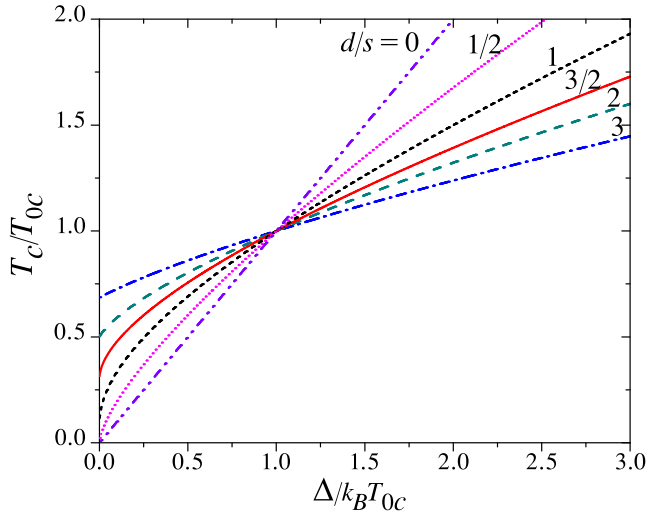}
\caption{(Color online) Critical temperature $T_c/T_{0c}$ as a function of $\Delta/k_BT_{0c}$ for values $d/s$ = 0, 1/2, 1, 3/2, 2 and 3.}
  \label{grafT0c_Tc_delta_variable}
 \end{figure}%
 \begin{figure}
  \centering
\hspace{1cm}
\includegraphics[scale=0.4]{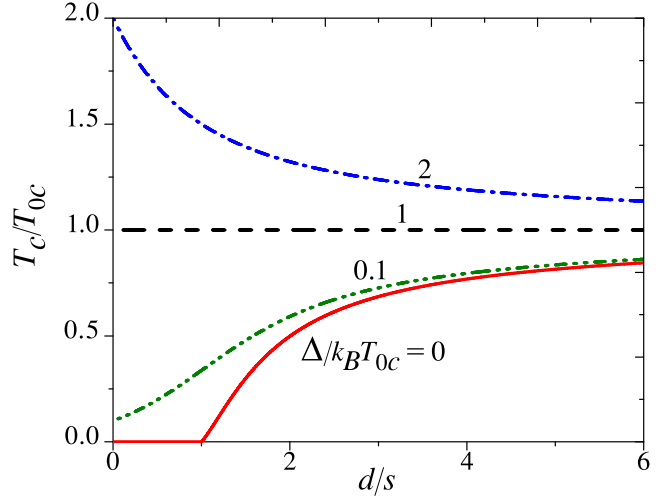}
\caption{(Color online) Critical temperature $T_c/T_{0c}$ as a function of $d/s$ for values $\Delta/k_BT_{0c}$ = 0, 0.1, 1 and 2.}
\label{graf:T0c_Tc_ds_variable}
\end{figure}


In Fig. \ref{grafT0c_Tc_delta_variable} we show the critical temperature in units of $T_{0c}$, which increases monotonically as a function of the energy gap $\Delta/k_BT_{0c}$. It is important to note that the critical temperature becomes zero if the gap is zero for values of $d/s \leq 1$, which means that without the energy gap there is no BEC at a temperature different from zero. For values $d/s > 1$ the BEC critical temperature is different from zero even for a zero energy gap.

In Fig. \ref{graf:T0c_Tc_ds_variable} we show the critical temperature as a function of $d/s$ for different values of $\Delta/k_BT_{0c}$. From the curves we note that the critical temperature $T_c$ is $T_{0c}$ as $d/s$ tends to infinity for any value of $\Delta/k_BT_{0c}$. This means that systems with $d\gg 1$ (and/or $s\ll 1$) the value of the critical temperature is independent of the energy gap.


\subsection{Condensate Fraction}
Dividing Eq. (\ref{num_bos_e}) by $N$ and solving for $N_0/N$ we have
\begin{equation}\label{eq:condensado_con_V}
N_0/N=1-\frac{T^{d/s}g_{\frac{d}{s}}(z_1)}{\left(\frac{c_s}{k_B}\right)^{d/s}\left(\frac{s\Gamma(d/2)(2\pi)^{d}n}{2\pi^{d/2}\Gamma(d/s)}\right)}
\end{equation}  
which can be rewritten in terms of the critical temperature $T_c$ (\ref{tem_critica}) in the following way
\begin{equation}\label{eq:condensado_para_todaT}
N_0/N=1-\left(\frac{T}{T_c}\right)^{d/s}\frac{g_{\frac{d}{s}}(z_1)}{g_{\frac{d}{s}}\left(e^{-\frac{\Delta}{k_BT_c}}\right)}.
\end{equation}
%
%
For $T<T_c$ the chemical potential is equal to the energy of the ground state which means that $z_1 \rightarrow e^{-\frac{\Delta}{k_BT}}$  and therefore the condensate fraction is
\begin{equation}\label{eqcondensado}
\frac{N_0}{N}=1-\left(T/T_c\right)^{d/s}\frac{g_{\frac{d}{s}}\left(e^{-\frac{\Delta}{k_BT}}\right)}{g_{\frac{d}{s}}\left(e^{-\frac{\Delta}{k_BT_c}}\right)}.
\end{equation}
When $d/s$ approaches zero the Bose function $g_{d/s}(z_1)$ becomes a geometrical series which can be changed by the quotient $z_1/(1-z_1)$, then the last expression becomes
\begin{equation}
\frac{N_0}{N}=\frac{1-\exp \left[-\frac{\Delta}{k_B}\left( \frac{1}{T}-\frac{1}{T_c}\right) \right]}{1-\exp \left[-\frac{\Delta}{k_BT} \right]}    \xrightarrow[d/s\rightarrow 0]{} 1.
\end{equation} 
Note that when $d/s\rightarrow0$
 the critical temperature $T_c$ goes to infinity and the condensate fraction goes to unit as $\Delta\neq0$.
\begin{figure}
\centering
\hspace{1cm}
  \includegraphics[scale=0.4]{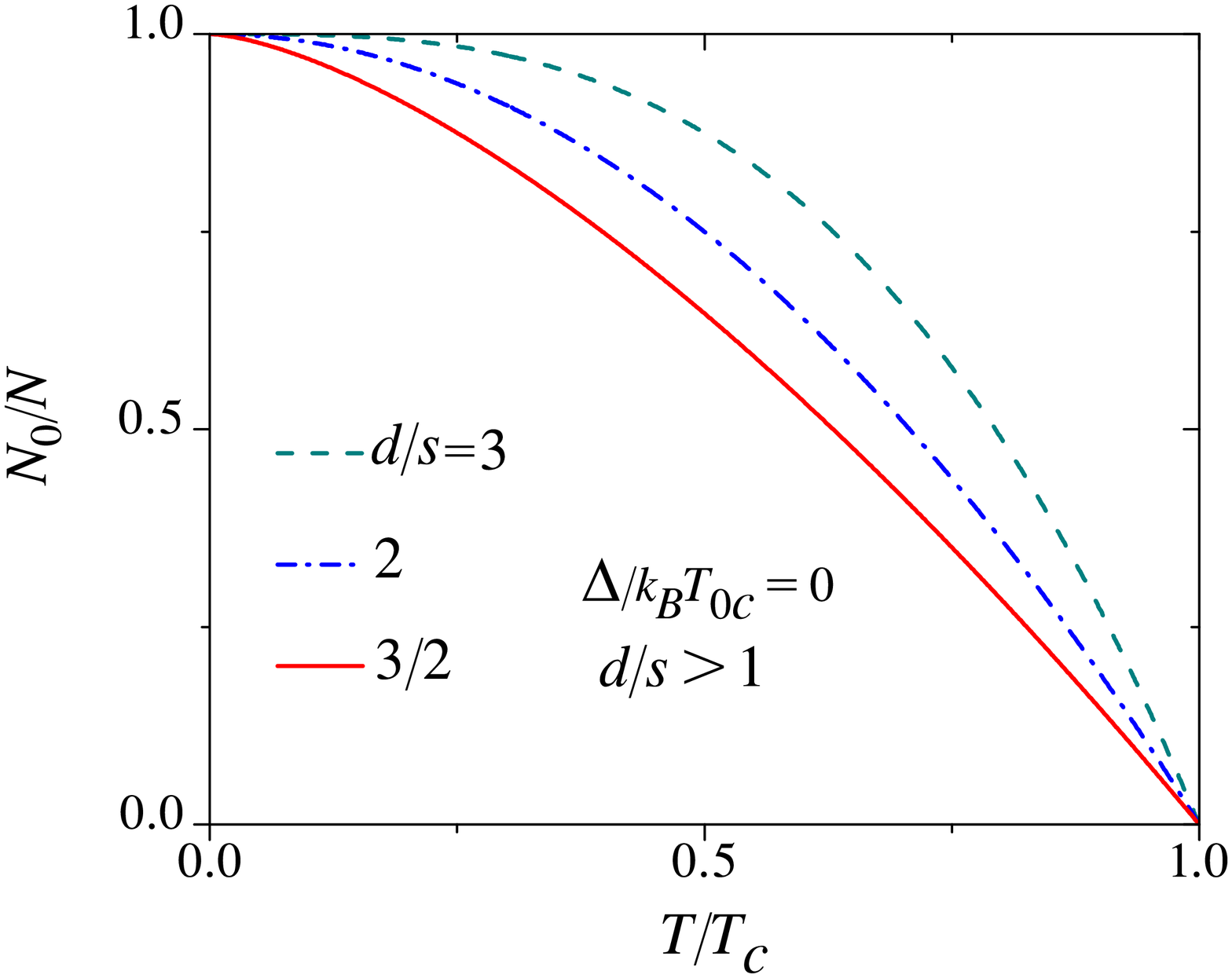}
 \caption{(Color online) Condensate fraction as a function of $T/T_c $ with $\Delta=0$, for $d/s$ = 3/2, 2 and 3.}
  \label{grafcondensado-delta-cero}
 \end{figure}%
 \begin{figure}
  \centering
  \hspace{1cm}
  \includegraphics[scale=0.4]{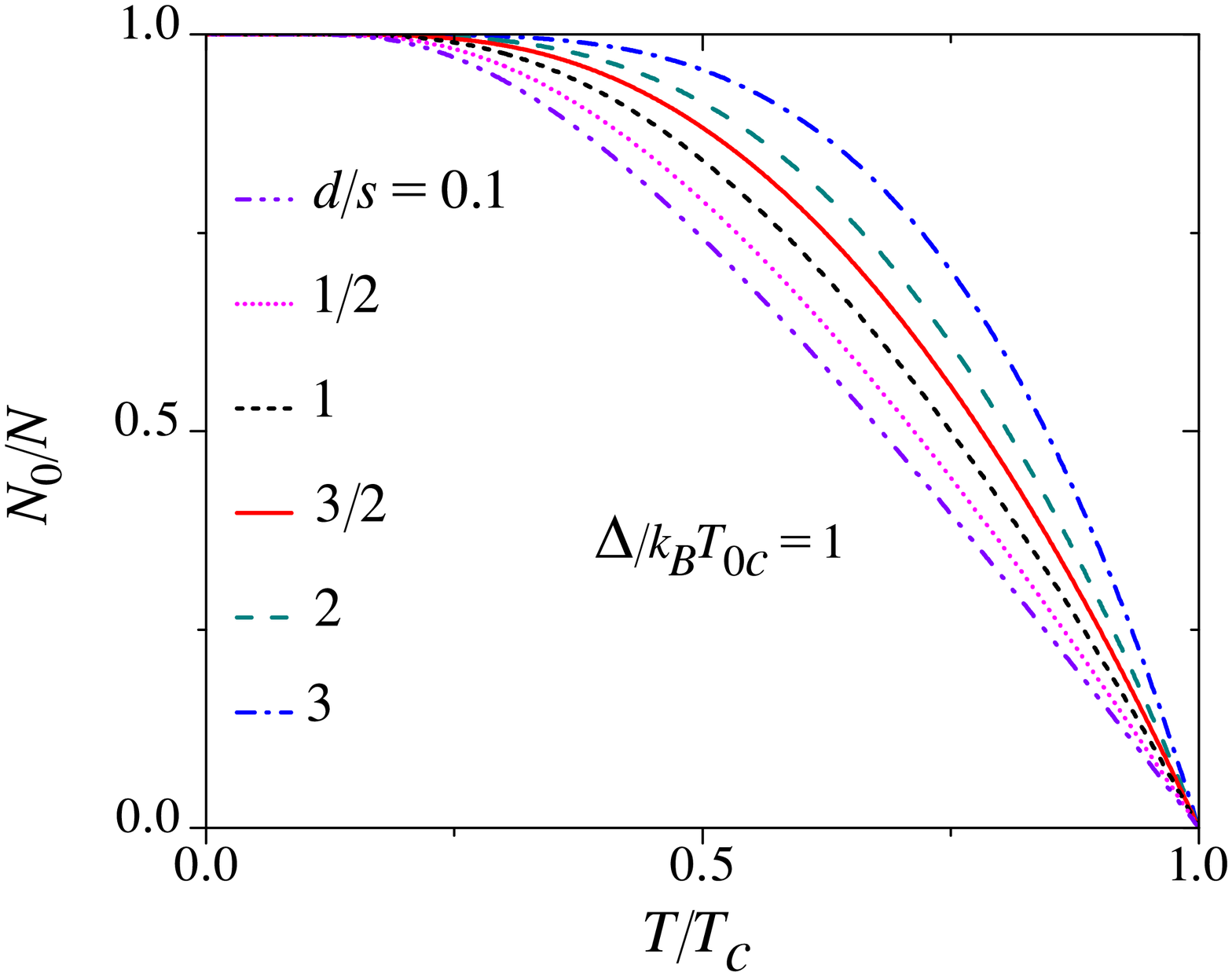}
\caption{(Color online) Condensate fraction as a function of $T/T_c $ with $\Delta=k_BT_c$ for $d/s$ = 0.1, 1/2, 1, 2, and 3. }
\label{grafcondensado-delta-1}
\end{figure}

In Figs. \ref{grafcondensado-delta-cero} and \ref{grafcondensado-delta-1} we show the behavior of the condensate fraction as a function of temperature $T/T_c$ in the interval [0,1]; in both cases, $\Delta = 0$ and $\Delta/k_BT_{0c} = 1$, the condensate fraction decreases monotonically. 
  Also, in both figures we show that the condensate fraction grows if $d/s$ grows for any $T/T_c \in [0,1] $. These results are consistent with the fact that $N_0/N = 1$ if $d$ tends to zero because in this limit the critical temperature tends to infinity, therefore the quotient $T/T_c$ is practically zero for any value of system temperature and the condensate fraction is one, as we can see in Fig. \ref{grafcondensado-delta-1} for small values of $T/T_c$.
\section{Thermodynamics properties}
\subsection{Equation of State}
In order to calculate the equation of state we start with the well known expression for the grand canonical function
\begin{equation}
\frac{PV}{k_BT}=-\sum_{\varepsilon}\ln \left(1-e^{-\beta(\varepsilon-\mu)}\right).
\end{equation}
By changing the sum by the integral over the $d$-dimensional sphere 
%
%
and after doing the integral we get 
\begin{equation}
\frac{PV}{Nk_BT}=-\frac{\ln\left(1-e^{-\beta(\varepsilon_0-\mu)}\right)}{N}+\left(\frac{T}{T_c}\right)^{d/s}\frac{g_{\frac{d}{s}+1}(z_1)}{g_{\frac{d}{s}}\left(e^{-\frac{\Delta}{k_BT_c}}\right)}.
\label{eq:PV1}
\end{equation}
%
%
%
The first term in the right member of (\ref{eq:PV1}) can be written in term of the number of bosons in the ground state, $N_0=1/\left(e^{\beta(\varepsilon_0-\mu)}-1\right)$, in the 
following way
%
%
\begin{equation}
-\ln\left(1-e^{-\beta(\varepsilon_0-\mu)}\right)/N=\ln(N_0+1)/N,
\end{equation}
therefore in the T.L., where $N\gg1$, this term is negligible for any temperature. For this reason the pressure is
\begin{equation}\label{eq:presion}
P=\frac{Nk_BT}{V}\left(\frac{T}{T_c}\right)^{d/s}\frac{g_{\frac{d}{s}+1}(z_1)}{g_{\frac{d}{s}}\left(e^{-\frac{\Delta}{k_BT_c}}\right)}.
\end{equation}
For temperatures $T \gg T_c$ we recover the classical limit (C.L.) $PV = Nk_BT$. However for $T < T_c$ the system behavior is strongly modulated by the presence of the gap.

\subsection{Generalized Thermal Wavelength}
Equation (\ref{eq:presion}) can be shortly written in terms of a generalized thermal de Broglie wavelength $\lambda$ as  
\begin{equation}
\frac{P}{k_BT}=\frac{g_{\frac{d}{s}+1}(z_1)}{\lambda^d}
\end{equation}
where 
\begin{equation}\label{thermwavelengthondelta}
\lambda=\left(\frac{V}{N}\right)^{1/d}\left(\frac{T_c}{T}\right)^{1/s} \left(g_{\frac{d}{s}}\left(e^{-\frac{\Delta}{k_BT_c}}\right)\right)^{1/d} 
\end{equation}
which after introducing the expression (\ref{tem_critica}), reduces to
\begin{equation}\label{thermwavelengthYan}
\lambda=2\sqrt{\pi} \left(\frac{c_s}{k_B T}\right)^{1/s} \left( \frac{\Gamma (d/2+1)}{\Gamma (d/s+1)} \right)^{1/d} .
\end{equation}
Last expression is equal to that  
obtained in Ref. \onlinecite{YanZEJP2000} for the case with $\Delta = 0$. Note that the generalized thermal wavelength (\ref{thermwavelengthYan}) is $\Delta$ independent and, additionally, for a quadratic dispersion relation $s=2$ we recover
$\lambda=2\sqrt{\pi c_2/(k_B T)}=h/\sqrt{2 \pi k_B T}$, which is also $d$ independent. 

\subsection{Internal Energy}
The internal energy of the system is given by the sum of the energies of each particle, including those in the ground state. As we have a gas of bosons then we can write the internal energy as 
\begin{equation}
U=N_0\varepsilon_0 + \sum_{\boldsymbol{k} \neq 0} \varepsilon_k n_k = N_0\varepsilon_0 + \sum_{\boldsymbol{k} \neq 0} \frac{\varepsilon_k}{e^{\beta(\varepsilon_k-\mu)}-1}.
\end{equation}
Approximating the sum by an integral, introducing the dispersion relation (\ref{eq:reldis}), carrying out
 the variable substitution $\varepsilon = \varepsilon-\Delta-\varepsilon_0$, rewriting the last expression in terms of the Bose function and carrying out some algebra we obtain
\begin{equation}
\begin{split}
U(L^d & ,T)=N_0\varepsilon_0 + \left(\frac{L}{2\pi}\right)^d \frac{2\pi^{\frac{d}{2}} \, \Gamma(d/s)}{sc_s^{\frac{d}{s}}\Gamma(d/2)}(k_B T)^{\frac{d}{s}} \times 
\\ &
 \left[\frac{d}{s}(k_B T) g_{\frac{d}{s}+1}(z_1)+(\varepsilon_0+\Delta)g_{\frac{d}{s}}(z_1)\right].
\end{split}
\end{equation}
%
%
%
%
From the last expression we can get the internal energy ratio between the  
 total internal energy minus the system ground state energy 
 $N\varepsilon_0$ and $Nk_BT$, which can be rewritten in terms of the critical temperature $T_c$ (\ref{tem_critica}) in the following way 
\begin{equation}\label{eqenergiaporparticula}
\frac{U-N\varepsilon_0}{Nk_BT}= \frac{(T/T_c)^{\frac{d}{s}}}{g_{\frac{d}{s}}\left(e^{-\frac{\Delta}{k_BT_c}}\right)}\left[\frac{d}{s} g_{\frac{d}{s}+1}(z_1)+\frac{\Delta}{k_BT} \, g_{\frac{d}{s}}(z_1)\right].
\end{equation} 
In terms of the internal energy the pressure expression \eqref{eq:presion} can be written as follows
\begin{equation}\label{eq:presionenergia}
\frac{d}{s}PV=U- N \left[ \varepsilon_0+\left(1-\frac{N_0}{N}\right)\Delta \right],
\end{equation}
which generalizes the Ec. (11) of \cite{Aguilera} to include an energy gap.
\begin{figure}[h!]
\centering
\includegraphics[scale=0.4]{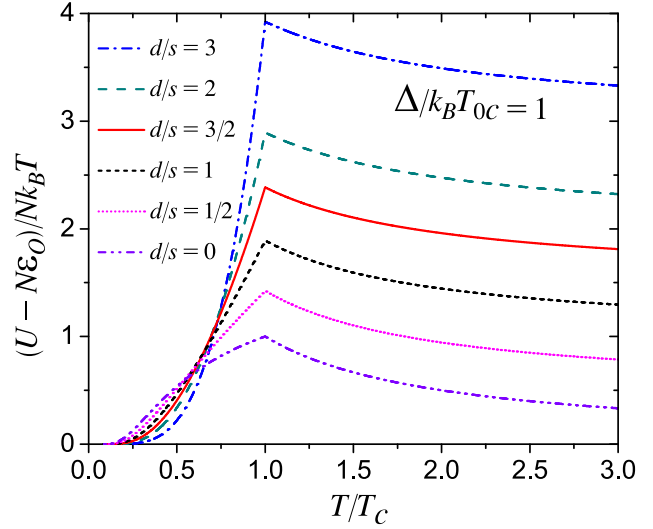}
\caption{(Color online) Internal energy ratio $(U-N\varepsilon_0)/(Nk_BT)$ as a function of the temperature $T/T_{0c}$.}
\label{graf:Energia_interna}
\end{figure}

In Fig. \ref{graf:Energia_interna} we plot the internal energy ratio as a function of temperature for several values of $d/s$.  For $T<T_{c}$ the internal energy ratio increases monotonically till it arrives to $T =T_{c}$ where it shows a peak which comes from the gap effect since it does not exist if $\Delta = 0$. For $T/T_{c}\gg1$ the internal energy referred to $N(\varepsilon_0+\Delta)$ satisfies $U-N(\varepsilon_0+\Delta) \rightarrow (d/s)Nk_BT$, which reproduces the expected C.L.
\subsection{Isochoric Specific Heat}
From the internal energy expression (\ref{eqenergiaporparticula}) we obtain the isochoric specific heat per particle, i.e.,
\begin{eqnarray}\label{eq:firstCv}
\frac{C_V}{Nk_B}=\frac{1}{Nk_B}\left(\frac{\partial U}{\partial T}\right)_{N,V}= 
\left(\frac{d}{s}\right)\left(\frac{T}{T_c}\right)^{d/s}\frac{1}{g_{\frac{d}{s}}\left(e^{-\frac{\Delta}{k_BT_c}}\right)} \times && \nonumber \\ \left[\left(1+\frac{d}{s}\right)g_{\frac{d}{s}+1}(z_1) \right. 
  +\left(\frac{\Delta}{k_BT}+T \left(\frac{\partial \ln(z_1)}{\partial T}\right)_{N,V} \right)g_{\frac{d}{s}}(z_1) && \nonumber \\
 \left. + T \left(\frac{s}{d}\right) \frac{\Delta}{k_BT} \left( \frac{\partial \ln(z_1)}{\partial T}\right)_{N,V}  g_{\frac{d}{s}-1}(z_1)\right]. \hspace{1.0cm} && 
\end{eqnarray}
As for $T > T_c$ the condensate fraction is negligible $N_0/N \simeq 0$ and differentiating 
\eqref{eq:condensado_para_todaT} with respect to $T$ we obtain
\begin{equation}\label{eqderivadadez_1}
\left( \frac{\partial \ln (z_1)}{\partial T}\right)_{N,V} = -\frac{d}{s}\frac{1}{T}\frac{g_{\frac{d}{s}}(z_1)}{g_{\frac{d}{s}-1}(z_1)}.
\end{equation}
Introducing \eqref{eqderivadadez_1} in \eqref{eq:firstCv} and using Eq. \eqref{eq:condensado_para_todaT} we arrive to the expression for the specific heat for $T > T_c$ as follows
%
%
%
\begin{equation}\label{eqcalorespecificoparat>t_c}
\begin{split}
\hspace{-0.30cm} \frac{C_V}{Nk_B}= \frac{d}{s} \left[  \left(1+\frac{d}{s}\right)\frac{g_{\frac{d}{s}+1}(z_1)}{g_{\frac{d}{s}}(z_1)} - \left(\frac{d}{s}\right) \frac{g_{\frac{d}{s}}(z_1)}{g_{\frac{d}{s}-1}(z_1)} \right].
\end{split}
\end{equation}
For $ T\gg T_c $, $g_{\frac{d}{s}+1}(z_1)/g_{\frac{d}{s}}(z_1) = g_{\frac{d}{s}}(z_1)/g_{\frac{d}{s}-1}(z_1) = 1$ and we recover the C.L. value $C_V/Nk_B = d/s$.
For $T\leq T_c$ the chemical potential takes the value of the ground state $\mu =\varepsilon_0$ and therefore the internal energy given by (\ref{eqenergiaporparticula}) can be rewritten as 
\begin{eqnarray}\label{eq:InternalEnergyLessT}
\frac{U-N\varepsilon_0}{N}= \frac{\left( T/T_c \right)^{d/s}}{g_{\frac{d}{s}}\left(e^{-\frac{\Delta}{k_BT_c}}\right)}\left[\frac{d}{s}(k_B T) g_{\frac{d}{s}+1}\left(e^{-\frac{\Delta}{k_BT}}\right) \right.  && \nonumber \\ 
\left. + \Delta \, g_{\frac{d}{s}}\left(e^{-\frac{\Delta}{k_BT}}\right)\right]. \hspace{0.70cm} &&
\end{eqnarray} 
Differentiating \eqref{eq:InternalEnergyLessT} with respect to $T$ we have that the specific heat for $T \leq T_c$ is
\begin{equation}\label{eqcalorespecificoparat<t_c}
\begin{split}
\frac{C_V}{Nk_B}= &
\frac{d}{s} \, \left( 1-\frac{N_0}{N}\right) \left[ \left(1+\frac{d}{s}\right)\frac{g_{\frac{d}{s}+1}\left(e^{-\frac{\Delta}{k_BT}}\right)}{g_{\frac{d}{s}}\left(e^{-\frac{\Delta}{k_BT}}\right)}  \right.
\\& \hspace{-1.0cm} + \left. \frac{s}{d}  \left(\frac{\Delta}{k_B T}\right) \left( \frac{d}{s} + \frac{\Delta}{k_B T} \frac{g_{\frac{d}{s}-1}\left(e^{-\frac{\Delta}{k_BT}}\right)}{g_{\frac{d}{s}}\left(e^{-\frac{\Delta}{k_BT}}\right)} \right) + \frac{\Delta}{k_B T} \right].
\end{split}
\end{equation}
We note that for $ T \ll \Delta /k_B$, i.e., $\Delta/k_BT\gg1$ the function $g_{\frac{d}{s}}\left(e^{-\frac{\Delta}{k_BT}}\right)\simeq e^{-\frac{\Delta}{k_BT}}$ for any value of $d/s$. Therefore the isochoric specific heat can be approximated by
\begin{eqnarray}
\frac{C_V}{Nk_B} \simeq \frac{d}{s}\frac{\left( T/T_c \right)^{d/s}e^{-\frac{\Delta}{k_BT}}}{g_{\frac{d}{s}}\left(e^{-\frac{\Delta}{k_BT_c}}\right)} \left[\left( 1+\frac{d}{s}\right) +\frac{s}{d}\left(\frac{\Delta}{k_BT}\right)^2 \right.&& \nonumber \\ 
\left. +\frac{2\Delta}{k_BT}\right].  \hspace{0.70cm} &&
\end{eqnarray}  
which generalize the Eq. (25a) given by London in Ref. \onlinecite{Londonb} p. 55. Then for very low temperatures the isochoric specific heat tends to zero exponentially as $e^{-\frac{\Delta}{k_BT}}$, since an exponential decays much faster than a polynomial.

Now in the limit when $d/s\rightarrow 0 $, Bose functions $g_{0}\left( e^{-\Delta/k_BT} \right) $ and $g_{-1} \left( e^{-\Delta/k_BT} \right) $ tend to $ e^{-\Delta /k_BT} / \left( 1 - e^{-\Delta / k_BT} \right) $ and $e^{-\Delta /k_BT} / \left( 1-e^{-\Delta/k_BT} \right)^2 $ respectively. 
Then from Eq. (\ref{eqcalorespecificoparat<t_c}) and using (\ref{eq:condensado_para_todaT}) 
\begin{eqnarray}
\frac{C_V}{Nk_B} \xrightarrow[d/s\rightarrow 0]{} \left(\frac{\Delta}{k_BT}\right)^2\frac{g_{-1}\left(e^{-\frac{\Delta}{k_BT}}\right)}{g_0 \left(e^{-\frac{\Delta}{k_BT_c}}\right)}= && \nonumber \\ \left(\frac{\Delta}{k_BT}\right)^2\frac{e^{\frac{\Delta}{k_BT_c}}-1}{\left(e^{\frac{\Delta}{k_BT}}-1\right)^2} e^{\frac{\Delta}{k_BT}}.&&
\end{eqnarray}   
\begin{figure}[hbt]
\centering
\hspace{1cm}
\includegraphics[scale=0.4]{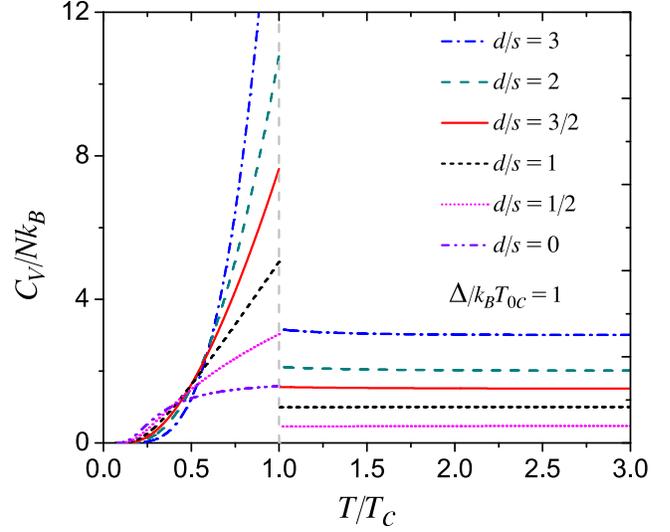}
\caption{(Color online) Isochoric specific heat as a function of $T/T_{0c}$, for $\Delta/k_BT_{0c}=1$ and for $d/s$ = 0, 1/2, 1, 3/2, 2, 3.}
\label{grafCalor_especifico}
\end{figure}
In Fig. \ref{grafCalor_especifico} we show the isochoric specific heat as a function of $T/T_{c}$, for $\Delta/k_BT_{0c}=1$ and several values of $d/s$. Every curve begins from zero and grows until it reaches its value at $T=T_{c}$, where they show a jump to another branch of smaller values, which in the C.L. tend to the value $d/s$.

\subsection{Isochoric Specific Heat Jump $\boldsymbol{\Delta} C_V$ }

The jump $\boldsymbol{\Delta} C_V$ in the isochoric specific heat at $T_c$ is obtained by subtracting the expression (\ref{eqcalorespecificoparat>t_c}) from (\ref{eqcalorespecificoparat<t_c}) both evaluated at $T=T_c$. After some algebra we have 
\begin{equation}\label{eq:saltocV}
\begin{split}
& \frac{\boldsymbol{\Delta} C_V}{Nk_B} \equiv \frac{C_V(T_c^-) - C_V(T_c^+)}{Nk_B}= 
\\ &  \frac{g_{\frac{d}{s}} \left( z_{0c} \right) }{g_{\frac{d}{s}-1}\left( z_{0c} \right)} \left[ \frac{d}{s} + \frac{\Delta}{k_BT_c} \frac{g_{\frac{d}{s}-1} \left( z_{0c} \right) }{g_{\frac{d}{s}}\left( z_{0c} \right)}\right]^2,
\end{split}
\end{equation}
where $z_{0c} \equiv e^{-\frac{\Delta}{k_BT_c}}$.

The last expression (\ref{eq:saltocV}) shows that for every $d/s > 0$ there will be a jump in the isochoric specific heat if $\Delta > 0$. Furthermore when $d/s \to 0$, ${\boldsymbol{\Delta} C_V}/{Nk_B}$ arrives to the constant $e/(e-1)$. 
\begin{figure}[h!]
\centering
\hspace{1cm}
\includegraphics[scale=0.4]{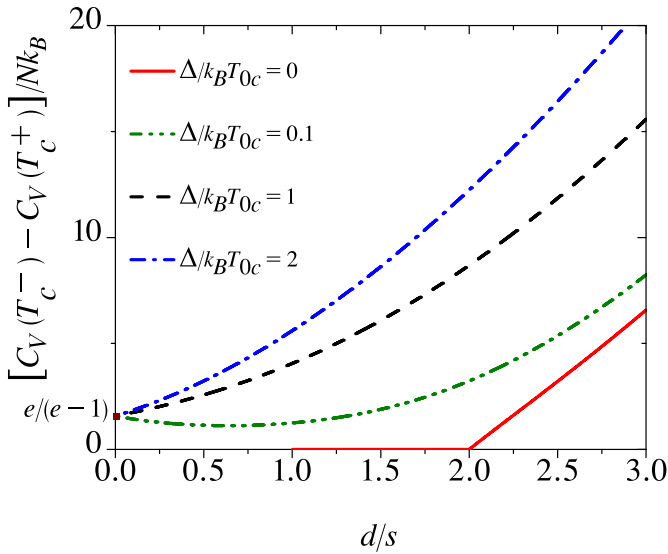}
\caption{(Color online) Isochoric specific heat jump at $T_c$ as a function of $d/s$, for $\Delta/k_BT_{0c}$ = 0, 0.1, 1 and 2. }
\label{graf:Salto_Cv_ds}
\end{figure}

\begin{figure}[hbt]
\centering
\hspace{1cm}
\includegraphics[scale=0.4]{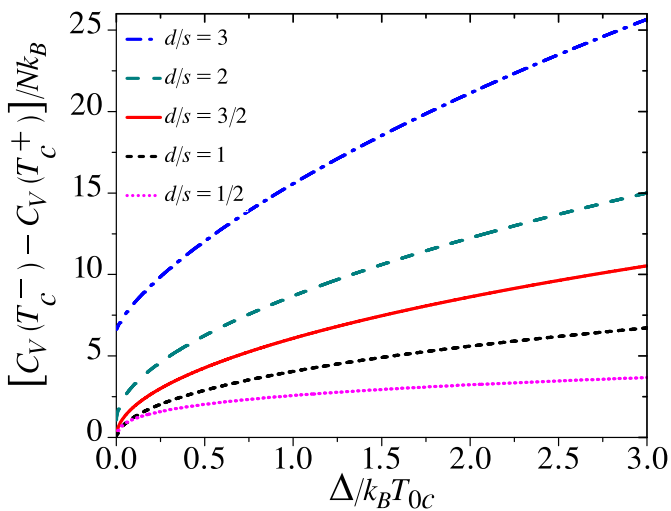}
\caption{(Color online) Isochoric specific heat jump at $T_c$ as a function of $\Delta/k_BT_{0c}$, for $d/s$ = 1/2, 1, 3/2, 2 and 3.}
\label{graf:Salto_Cv_delta}
\end{figure}

In Fig. \ref{graf:Salto_Cv_ds} we show the isochoric specific heat jump at $T_c$ as a function of $d/s$, for $\Delta/k_BT_{0c} = 1$ and 0. If $\Delta=0$, the jump is different from zero if and only if $d/s > 2$, as reported in Ref. \onlinecite{Aguilera}.
For $\Delta\neq 0$ there is a jump in the specific heat for $d/s > 0$ that grows monotonically as a function of $d/s$ for $\Delta/k_BT_{0c}=$1, 2, but for $\Delta/k_BT_{0c}=0.1$ the jump decreases for small values of $d/s$, presenting a minimum near $d/s=0.75$ and then increases monotonically as $d/s$ grows. For any value of $\Delta/k_BT_{0c}$ the isochoric specific heat jump is $e/(e-1)$ as $d/s$ tends to zero. In Fig. \ref{graf:Salto_Cv_delta} we show that the isochoric specific heat jump is a monotonically increasing as a function of $\Delta/k_BT_{0c}$.

\section{Conclusions}
	We have calculated the Bose-Einstein critical temperature, the condensate fraction, the equation of state, the internal energy, the ishocoric specific heat and the jump of ishocoric specific heat of a $d$-dimensional infinite Bose gas of permanent particles of mass $m$ and spin zero whose energy is proportional to its momentum to the power $s>0$ plus an energy gap between the ground and the first excited state energies as shown in Eq. (\ref{eq:reldis}). In summary, the existence of an energy gap between the ground and the first excited state energies produces a Bose-Einstein condensation at finite and non-zero temperature for any $d/s > 0$. On the other hand, if there is no gap, a $T_c \neq 0$ exists only if $d/s>1$, which 
	is smaller than the one with a gap what confirms that the presence of an energy gap increases the critical temperature for any $d/s>1$. For any temperature inside the interval [0, $T_c$] and for all $\Delta\geq 0$, the condensed fraction grows as $d/s$ grows, but without exceeding one.    

In the presence of an energy gap the calculated thermodynamic properties show the following behavior: the equation of state becomes $d$, $s$ and $\Delta$-dependent from where we deduce a $\Delta$ independent generalized thermal de Broglie wavelength, which generalize that one given in Ref. \onlinecite{YanZEJP2000}; for all $d/s>0$ the internal energy per particle has a peak at the critical temperature and, as a consequence the isochoric specific heat shows a jump at $T_c$, which differs from the case without a gap where the peak in the internal energy and the jump in the isochoric specific heat appear only for $d/s>2$. The magnitude of the heat capacity jump increases as $d/s$ grows. At very low temperatures the magnitude of $C_V$ grows exponentially as a function of temperature.  
We note that the ground state energy magnitude $\varepsilon_0$ has no effect on the behavior of thermodynamic properties except for increasing the internal energy magnitude per particle by the same amount. 
Finally, we show that our general relations for the critical temperature and thermodynamic properties reduce to well known cases for $d/s > 0$ and $\Delta=0$, in particular for a three-dimensional $d=3$ ideal Bose gas with $s=2$. 
 
\noindent {\bf Acknowledgments}. We acknowledge partial support from Grants PAPIIT-DGAPA-UNAM IN-107616 and CONACyT 221030.


\end{document}